\documentclass[12pt]{article}
\catcode`\@=11
\def\@normalsize{\@setsize\normalsize{15pt}\xiipt\@xiipt
\abovedisplayskip 14pt plus3pt minus3pt%
\belowdisplayskip \abovedisplayskip
\abovedisplayshortskip  \z@ plus3pt%
\belowdisplayshortskip  7pt plus3.5pt minus0pt}
\def\small{\@setsize\small{13.6pt}\xipt\@xipt
\abovedisplayskip 13pt plus3pt minus3pt%
\belowdisplayskip \abovedisplayskip
\abovedisplayshortskip  \z@ plus3pt%
\belowdisplayshortskip  7pt plus3.5pt minus0pt
\def\@listi{\parsep 4.5pt plus 2pt minus 1pt
            \itemsep \parsep
            \topsep 9pt plus 3pt minus 3pt}}
\def\underline#1{\relax\ifmmode\@@underline#1\else
        $\@@underline{\hbox{#1}}$\relax\fi}
\@twosidetrue
\relax
\catcode`@=12
\catcode`\@=11

\def\section{\@startsection{section}{1}{\z@}{3.5ex plus 1ex minus
   .2ex}{2.3ex plus .2ex}{\large\bf}}

\def\ps@headings{\def\@oddfoot{}\def\@evenfoot{}
\def\@oddhead{\hbox{}\hfill
        \makebox[.5\textwidth]{\raggedright\ignorespaces --\thepage{}--
        \hfill }}
\def\@evenhead{\@oddhead}
\def\subsectionmark##1{\markboth{##1}{}}
}
\ps@headings
\catcode`\@=12
\relax

\def\figcap{\section*{Figure Captions\markboth
        {FIGURECAPTIONS}{FIGURECAPTIONS}}\list
        {Fig. \arabic{enumi}:\hfill}{\settowidth\labelwidth{Fig. 999:}
        \leftmargin\labelwidth
        \advance\leftmargin\labelsep\usecounter{enumi}}}
 \relax
\def\tablecap{\section*{Table Captions\markboth
        {TABLECAPTIONS}{TABLECAPTIONS}}\list
        {Table \arabic{enumi}:\hfill}{\settowidth\labelwidth{Table 999:}
        \leftmargin\labelwidth
        \advance\leftmargin\labelsep\usecounter{enumi}}}
 \relax
\def\reflist{\section*{References\markboth
        {REFLIST}{REFLIST}}\list
        {[\arabic{enumi}]\hfill}{\settowidth\labelwidth{[999]}
        \leftmargin\labelwidth
        \advance\leftmargin\labelsep\usecounter{enumi}}}
 \relax
\catcode`\@=11
\def\marginnote#1{}
\newcount\hour
\newcount\minute
\newtoks\amorpm
\hour=\time\divide\hour by60
\minute=\time{\multiply\hour by60 \global\advance\minute by-
\hour}
\edef\standardtime{{\ifnum\hour<12 \global\amorpm={am}%
    \else\global\amorpm={pm}\advance\hour by-12 \fi
    \ifnum\hour=0 \hour=12 \fi
    \number\hour:\ifnum\minute<100\fi\number\minute\the\amorpm}}
\edef\militarytime{\number\hour:\ifnum\minute<100\fi\number\minute}

\def\draftlabel#1{{\@bsphack\if@filesw {\let\thepage\relax
  \xdef\@gtempa{\write\@auxout{\string
    \newlabel{#1}{{\@currentlabel}{\thepage}}}}}\@gtempa
    \if@nobreak \ifvmode\nobreak\fi\fi\fi\@esphack}
     \gdef\@eqnlabel{#1}}
\def\@eqnlabel{}
\def\@vacuum{}
\def\draftmarginnote#1{\marginpar{\raggedright\scriptsize\tt#1}}
\def\draft{\oddsidemargin -.5truein
        \def\@oddfoot{\sl preliminary draft \hfil
        \rm\thepage\hfil\sl\today\quad\militarytime}
        \let\@evenfoot\@oddfoot \overfullrule 3pt
        \let\label=\draftlabel
        \let\marginnote=\draftmarginnote
\def\@eqnnum{(\theequation)\rlap{\kern\marginparsep\tt\@eqnlabel}%
\global\let\@eqnlabel\@vacuum}  }
\def\preprint{\twocolumn\sloppy\flushbottom\parindent 1em
        \leftmargini 2em\leftmarginv .5em\leftmarginvi .5em
        \oddsidemargin -.5in    \evensidemargin -.5in
        \columnsep 15mm \footheight 0pt
        \textwidth 250mmin      \topmargin  -.4in
        \headheight 12pt \topskip .4in
        \textheight 175mm
        \footskip 0pt
\def\@oddhead{\thepage\hfil\addtocounter{page}{1}\thepage}
        \let\@evenhead\@oddhead \def\@oddfoot{} \def\@evenfoot{}
}
\def\titlepage{\@restonecolfalse\if@twocolumn\@restonecoltrue\onecolumn
     \else \newpage \fi \thispagestyle{empty}\c@page\z@
        \def\thefootnote{\fnsymbol{footnote}} }
\def\endtitlepage{\if@restonecol\twocolumn \else  \fi
        \def\thefootnote{\arabic{footnote}}
        \setcounter{footnote}{0}}  
\catcode`@=12
\relax

\def\ps@headings{\def\@oddfoot{}\def\@evenfoot{}
\def\@oddhead{\hbox{}\hfill
        \makebox[.5\textwidth]{\raggedright\ignorespaces --\thepage{}--
        \hfill }}
\def\@evenhead{\@oddhead}
\def\subsectionmark##1{\markboth{##1}{}}
}
\ps@headings
\relax

\newcommand{\newc}{\newcommand}
\newcommand{\fb}[0]{\overline{f}}
\newcommand{\hb}[0]{\overline{h}}
\renewcommand{\a}[1]{\alpha_{#1}}
\newc{\ra}{\rightarrow}
\newc{\lra}{\leftrightarrow}
\newc{\beq}{\begin{equation}}
\newc{\be}{\begin{equation}}
\newc{\eeq}{\end{equation}}
\newc{\ee}{\end{equation}}
\newc{\bea}{\begin{eqnarray}}
\newc{\eea}{\end{eqnarray}}

\newc{\sm}{Standard Model}
\newc{\smd}{Standard Model}
\newc{\ba}{\begin{eqnarray}}
 \newc{\ea}{\end{eqnarray}}
\newcommand{\e}{\varepsilon}
\newcommand{\eb}{\bar\varepsilon}
\newcommand{\kb}{\bar k}
\renewcommand{\l}{\lambda}

\begin{document}
\def\firstpage#1#2#3#4#5#6{
\begin{titlepage}
\nopagebreak
\title{\begin{flushright}
        \vspace*{-0.8in}
{ \normalsize  hep-ph/0002263 \\
CERN-TH/2000-064 \\
  IOA-02/2000 \\
February 2000 \\
}
\end{flushright}
\vfill
{#3}}
\author{\large #4 \\[1.0cm] #5}
\maketitle
\vskip -7mm
\nopagebreak
\begin{abstract}
{\noindent #6}
\end{abstract}
\vfill
\begin{flushleft}
\rule{16.1cm}{0.2mm}\\[-3mm]

\end{flushleft}
\thispagestyle{empty}
\end{titlepage}}
 \def\simlt{\stackrel{<}{{}_\sim}}
\def\simgt{\stackrel{>}{{}_\sim}}
\date{}
\firstpage{3118}{IC/95/34}
{\large\bf
Implications of Anomalous $U(1)$ Symmetry in Unified Models:
the Flipped $SU(5) \times U(1)$ Paradigm}
{{John Ellis}$^{\,a}$, {G.K. Leontaris}$^{\,b}$ and
{J. Rizos}$^{\,b}$}
{\normalsize\sl
$^a$Theory Division, CERN, CH 1211 Geneva 23, Switzerland\\[2.5mm]
\normalsize\sl
$^b$Theoretical Physics Division, Ioannina University,
GR-45110 Ioannina, Greece\\[2.5mm]
 }
{
 A generic feature of string-derived models is the
appearance of an anomalous Abelian $U(1)_A$ symmetry which,
among other properties, constrains the Yukawa couplings and
distinguishes the three families from each other. In this paper,
we discuss in a model-independent way
the general constraints imposed by such a $U(1)_A$
symmetry on fermion masses, $R$-violating couplings and
proton-decay operators in a generic flipped $SU(5) \times U(1)'$ model.
We construct all possible viable
fermion mass textures and give various examples of effective 
low-energy models which are distinguished from each other by their
different predictions for $B$-, $L$- and $R$-violating
effects. We pay particular attention to predictions for neutrino masses,
in the light of the recent Super-Kamiokande data.}

\newpage

\section{Introduction}

The minimal supersymmetric extension of the Standard Model (MSSM) is a
popular extension of the Standard Model (SM) of the electroweak and
strong interactions. The MSSM is the simplest extension of the SM which
solves the naturalness aspect of the hierarchy problem. Moreover, it
predicts cold dark matter with a density that could well be in the range
favoured by astrophysical observations~\cite{EHNOS}, and a relatively
light Higgs
boson, as hinted by the precision electroweak data from LEP and
elsewhere~\cite{LEPEWWG}.
Moreover, several supersymmetric grand unified (SUSY GUT) or partially
unified
extensions of the MSSM, based on gauge groups like $SU(4)$, $SU(5)$,
$SO(10)$, etc.~\cite{unified}, provide successful predictions of the
electroweak mixing angle. This is based on an appealing unification
of the strong and electroweak forces at an energy scale of the order of
$10^{16}$~GeV. Such SUSY GUT extensions of the Standard Model also
suggest
the possibility of unifying the above three forces with gravity in the
context of a supergravity or superstring scenario close to the Planck
scale. Since this picture provides a rather promising
framework for a more complete theory, we think it appropriate to
take a step further by analyzing in more detail
various additional phenomenological questions.

Among the large number of questions to be
answered in a more complete model of
elementary-particle interactions
is the origin and the strength of the Yukawa couplings.
As is well known, the fermion masses arise in the MSSM
from the Yukawa couplings in the superpotential through the vevs
of the two Higgs doublets $h_d,h_u$ which couple to quarks and leptons
to form mass terms as follows:
\ba
{\cal W} &=& \lambda_{ij}^u q_i u^c_j h_u + \lambda_{ij}^dq_i d^c_j h_d
               +\lambda_{ij}^e \ell_{i} e^c_j h_d.
\label{sup1}
\ea
There are several basic questions about the nature of the above
couplings which one may address: \\ (1) Although all these terms
are invariant under the (MS)SM gauge group, there is no
explanation why some of the Yukawa couplings are
much smaller than others, as is required in order to account for the
fermion mass hierarchy. \\
(2) In addition to the standard Yukawa couplings
which provide masses for the quarks and leptons, the gauge symmetry
and supersymmetry of the MSSM also allow terms which violate baryon
and lepton number already at the renormalizable level, namely:
\ba {\cal
W}&=&\lambda_{ijk}\ell_i\ell_je^c_k + \lambda'_{ijk}\ell_iq_jd^c_k
+\lambda_{ijk}''u^c_id^c_jd^c_k +\mu_i\ell_i h_u
\label{sup2}
\ea
It would therefore be natural to allow also
their existence in the Yukawa Lagrangian. In order, however, to
avoid rapid baryon decay induced by
these interactions, any combination of the $\lambda'$ and
$\lambda''$ couplings would have to be extremely
small.\\
(3) If the MSSM is to be embedded in a more complete
theory which includes gravity, non-renormalizable terms
will be induced, suppressed by powers of the Planck mass or some related
scale. Some of these
terms would have important observational effects, such as the $QQQ\ell$
operator, which cannot be eliminated by postulating conservation
of $R$ parity~\cite{Ibanez:1992pr}.

The most appropriate strategy for allowing some of the Yukawa
couplings in the Lagrangian while forbidding others is to
postulate additional symmetries. In particular, with regard to the
problem (2) mentioned above, the unwanted baryon- and
lepton-number-violating interactions given by the first three
terms of (\ref{sup2}) would be prevented by the matter parity,
known as $R$ parity~\cite{Farrar:1978xj}, which is defined as \ba
R=(-1)^{(3 B+L+2 S)} \ea where $B,L$ and $S$ are the baryon,
lepton and spin quantum numbers. Under this symmetry, all SM
particles are even, whilst their superpartners are odd. Imposing
this $R$-parity symmetry, all dangerous  terms in (\ref{sup2})
change sign and are eliminated from the superpotential, whilst the
Higgs-mixing $\mu$ term \ba \mu h_uh_d \ea has even $R$ parity,
and is allowed among the Yukawa couplings in the MSSM
Lagrangian~\footnote{The existence of this term is
phenomenologically necessary, with the mass parameter $\mu$ being
of the order of the electroweak scale. This is potentially an
issue, since in many models the $\mu$ term is generated
dynamically via the vev of some scalar field(s), and one must ask
whether the electroweak scale is a natural order of magnitude.}.


Another possibility, however, is that $R$ violation does occur
at some level, which would have interesting
phenomenological
implications. Some at least of the couplings in (\ref{sup2})
could be non-zero, although they should be adequately suppressed.
Stringent bounds on products of couplings arise from various
exotic reactions, some of them presented for convenience in
Table~\ref{Rbounds}.
\begin{table}
\begin{center}
\begin{tabular}{|c|c|}
\hline
Constrained $R$ Couplings &Reaction\\\hline
$\l_{1k1}\l_{1k2}$,$ \l_{231}\l_{131}< 7\times 10^{-7}$&$\mu\ra 3 e$\\
$\l'_{1k1}\l'_{2k1},\l'_{11k}\l'_{21k}<5\times 10^{-8}$&$\mu\;Ti\ra e \;Ti$\\
$\l'_{1k1}\l'_{2k2} < 8\times 10^{-7}$&$K_L\ra\mu e$\\
$\l'_{i12}\l'_{i21} < 10^{-9}$&$\Delta m_K$\\
$\l'_{k13}\l'_{k31} < 8 \times 10^{-8}$&$\Delta m_B$\\
${\rm Im}[\l'_{k12}\l^{\prime*}_{k21}] < 8\times 10^{-12}$&$\epsilon_K$\\
$\lambda'_{11k}\lambda''_{11k}<10^{-24}, k=2,3$&$p\ra \pi^+ 
(K^+)\bar{\nu}$\\
$\lambda'_{ijk}\lambda''_{\ell mn}<10^{-9}$&$p\ra \pi^+ (K^+)\bar{\nu}$\\
$\lambda''_{112}\lambda_{\ell33,3m3}<10^{-21}$&$p\ra K^+\bar{\nu}$\\
$\lambda''_{112}\lambda_{\ell22,2m2}<10^{-20}$&$p\ra K^+\bar{\nu}$\\
$\lambda''_{112}\lambda_{\ell11,1m1}<10^{-17}$&$p\ra K^+\bar{\nu}$\\
$\lambda''_{112}\lambda_{123,213}<10^{-14}$&$p\ra K^++3{\nu}$\\
$\lambda''_{112}\lambda_{132,312,231,321}<10^{-16}$&$p\ra
K^++e^\pm+\mu^{\mp}+{\nu}$\\
\hline
\end{tabular}
\end{center}
\caption{\label{Rbounds}
{\it Upper bounds on various  $\l_{ijk},\l'_{lmn}$ and $\l''_{opq}$
$R$-violating couplings from rare
processes~\cite{BP}. Limits on the products of pairs
of couplings are scaled by $\left(\tilde{m}/(100 GeV)\right)^2$,
where $\tilde{m}$ is the mass of the relevant supersymmetric particle
exchanged}.}
\end{table}

In order to obtain a consistent extension of the low-energy
phenomenological
Lagrangian, the MSSM gauge group $SU(3) \times SU(2) \times U(1)$
should be embedded in a higher symmetry.
Non-Abelian GUT groups such as $SO(10)$ and $SU(5)$ provide some
successful
relations between the various Yukawa couplings~\cite{CEG}, but cannot
eliminate
all  the unwanted  terms in (\ref{sup2}).  An important   role
in resolving these issues can be played by additional $U(1)$
symmetries, which were
originally used to  solve the fermion mass hierarchy problem~\cite{fn}.

One consistent picture which may give a final answer to these questions
has emerged in
effective field theory models  derived from strings.  In these
constructions,
one usually encounters a unified or partially unified  non-Abelian
gauge symmetry,  accompanied by a number of external $U(1)$ factors,
which may play the role of family symmetries~\footnote{See, for example,
\cite{a,b,c} and references therein.}. The fermion mass
textures and other Yukawa interactions
of the above models are dictated by the particular charges of
the particles under the $U(1)$ symmetries, the specific flat
direction which has been chosen, and further string selection rules
and other string symmetries. Several attempts have had successful
predictions: to our knowledge,
the most promising results have been obtained in
the context of the free-fermionic
formulation of the heterotic superstring~\cite{Antoniadis:1987rn}.
To construct a model, one has first to determine a suitable set of
boundary conditions on the world-sheet fermionic degrees of freedom
introduced to cancel the conformal anomaly.
In this context, all properties of the effective field theory model,
i.e., the gauge symmetry group including external $U(1)$ factors, other
continuous or discrete symmetries, the Yukawa interactions and
their strengths are in principle  determined once the flat directions
are specified.

Although this approach looks very promising,
there are  shortcomings, which have mainly to do with the freedom
of choice in the string vacuum. There are
many possible choices for the sets of boundary conditions
on the basis vectors, and each one of them leads to a different group
structure.  Moreover, even if the gauge group is specified
by some other considerations,  there are still many basis choices
that predict different massless spectra and Yukawa interactions.
A third source of ambiguity arises from the non-uniqueness
of the vacuum, associated with different flat directions of the
effective potential. These arise because a
string model typically contains, in addition to the MSSM spectrum,
a number of singlet fields with zero electric charge. These singlets
may acquire non-zero vevs and provide masses to various
fermions and other particles through their tree-level and non-renormalizable
Yukawa couplings. Their vevs have to respect certain constraints
necessary to ensure the $D$- and $F$-flatness of the superpotential.
In general, there are multiple solutions of these flatness conditions,
and one has to choose the most appropriate of them guided
by phenomenological criteria. In other words, even within the class of
models sharing a
given string basis, there are multiple string vacua, and
it is not known {\it a priori} which is the most suitable.

It is interesting to note, however, that most of
the known string models share several generic properties~\cite{a,b,c}
which  point to an one particular generalization of the
MSSM. Two of the most important are the existence (i) of an
anomalous $U(1)_A$ symmetry and (ii) various singlets capable of
obtaining non-zero vevs, whose magnitude
is fixed once the string vacuum is chosen.  Based on the above observations,
during the last few years there has developed a new
strategy for attacking the problems of fermion masses and other Yukawa
interactions. Extensions of the MSSM with  new Abelian or discrete
symmetries~\cite{fn,Ibanez:1993fy,Ibanez:1994ig,Leurer:1994gy,
Binetruy:1995ru,Nir,Dreiner:1995ra,Dudas:1995yu,LR,GRU1} have given
some insights into the fermion mass hierarchy puzzle thanks to
anomalous or non-anomalous $U(1)$ symmetries which distinguish
various families. It has been established that, in the MSSM, the
existence of one non-anomalous $U(1)$ family symmetry is
sufficient to obtain a viable hierarchical fermion mass
pattern. Additional non-anomalous $U(1)$ symmetries may also be
possible, resulting in further restrictions on the Yukawa
Lagrangian. Interestingly, the appearance of multiple $U(1)$
symmetries is generic in models derived from the superstring.

In this work, we adopt a slightly different point of view, and go
beyond the simple $U(1)_A$ extensions of the MSSM. As described
above, another important ingredient of string-derived models
is the existence of some intermediate gauge
symmetry~\footnote{For earlier attempts, see
also~\cite{ALL,GRU1}.}, instead of the Standard Model one. As found
previously, in most cases this intermediate gauge 
symmetry provides successful
relations between the Yukawa couplings. Moreover, it is known that
there are only a few types of such partially-unified groups which
dispense
with the use of Higgs fields in the adjoint representation~\cite{e} to
break down to the Standard Model symmetry. Here we provide a general
discussion of anomalous $U(1)_A$ family symmetries
within the context of the flipped $SU(5)\times U(1)'$ gauge group.
Our {\it first} aim is to determine the possible $U(1)_A$
family symmetries which predict fermion mass matrices consistent
with the low-energy measurements. This and the specific
non-Abelian structure of the theory impose conditions on the
$U(1)_A$ charges of the matter and Higgs representations. At a
{\it second} stage, we explore the role of the $U(1)_A$ symmetries in
constraining
$R$-violating interactions, baryon-number-violating operators,
and Higgs mixing terms.  These constraints further reduce the
acceptable
choices of $U(1)_A$ charges, and therefore the possible fermion
mass textures. {\it Thirdly}, an
interesting feature of flipped $SU(5)\times
U(1)'$, as opposed to the minimal $SU(5)$ GUT, is that the neutrinos
obtain naturally non-zero masses.
In the type of model discussed here, we find very restrictive
forms for the neutrino mass matrices.

Section 2 is devoted to the
presentation of the minimal supersymmetric version of flipped
$SU(5)\times U(1)'$. In Section 3 we derive the possible forms of $R$-,
$B$- and $L$-violating operators in $SU(5)\times U(1)'$,
and relate them to the
$\lambda,\lambda'$ and $\lambda''$ parameters of (2). In Section 4
we discuss the anomaly cancellation conditions in conjunction with
the canonical $\sin^2\theta_W$ condition at the unification scale. The
charged fermion mass textures and the general solutions of the
anomaly-cancellation conditions determining the $U(1)_A$ charges
are given in Section 5. Section 6 deals with the neutrino masses
and the $B$-, $L$- and $R$-violating couplings in the presence
of the anomalous $U(1)_A$ symmetry. Specific detailed models are
presented and discused in Section 7, 
and future extensions and prospects are presented
in Section 8.

\section{Review of the Flipped $SU(5)\times U(1)'$ Model}

We present in this Section basic features of the supersymmetric
version of the flipped $SU(5)\times U(1)'$ model. For later purposes and
for comparison with the Standard Model charge assignments, it
is useful to recall the decomposition of the $SU(5)\times U(1)'$
representations in terms of the Standard Model spectrum, which are
presented in Table~\ref{decom}.
\begin{table}
\begin{center}
\begin{tabular}{|c|c|}
\hline
$SU(5)\times U(1)$&$SU(3)\times SU(2)\times U(1)$
\\
\hline
$F_i\left({10},\frac{1}{2}\right)$&
$Q(3,2,\frac 16)+d^c(\bar{3},1,\frac 13)+\nu^c(1,1,0)$\\
$\bar f_i\left(\bar{5},-\frac{3}{2}\right)$&
$u^c(\bar{3},1,-\frac{2}{3})+\ell(1,2,-\frac 12)$\\
$e^c_i\left({1},\frac{5}{2}\right)$&
$e^c(1,1,1)$\\
$H\left({10},\frac{1}{2}\right)$&
$Q_H(3,2,\frac 16)+d^c_H(\bar{3},1,\frac 13)+\nu^c_H(1,1,0)$\\
$\bar H\left(\overline{10},-\frac{1}{2}\right)$&
$\bar Q_H(3,2,-\frac 16)+\bar d^c_H(\bar{3},1,-\frac 13)+\bar \nu^c_H(1,1,0)$\\
$h\left({5},-1\right)$&
$D(\bar{3},1,-\frac 13)+h_d(2,1,-\frac 12)$\\
$\bar h\left(\bar{5},1\right)$&
$\bar D(\bar{3},1,\frac 13)+h_u(2,1,\frac 12)$\\
\hline
\end{tabular}
\end{center}
\caption{\label{decom}{\it Decomposition of the $SU(5)\times U(1)'$
matter
and Higgs into representations of the Standard Model gauge group}.}
\end{table}
This is the minimal version, whose field content includes the
MSSM spectrum
augmented only by right-handed neutrinos and just two pairs of
colored triplets,
($d^c_H,D$) and $(\bar d^c_H,\bar D)$, which acquire high
masses.
The superpotential  of the minimal model contains the
following Yukawa couplings:
\ba
{\cal W} &=& \lambda_u F\bar{f}\bar{h} + \lambda_d F F h +
\lambda_{e}\ell^c\bar{f}h
\nonumber         \\
         &+&     \lambda_{\nu}F \bar{H}\phi_i
         + H H h + \bar{H} \bar{H} \bar{h} + \phi_i \phi_j \phi_k
+h\bar{h}\phi.
\label{su5sup}
\ea
The first line gives masses to the charged fermion fields of the Standard
Model, plus Dirac masses for the neutrinos. Four singlet fields, $\phi$
and $\phi_{1,2,3}$ are also introduced, but only one  of them
develops a non-zero vev:
$\langle \phi \rangle \approx m_W$.
The coupling $\lambda_{\nu}F \bar{H}\phi_i$
together with the Dirac mass terms obtained from the $F \bar f\bar h$
term realizes an extended see-saw mechanism which suppresses
the left-handed neutrino masses\cite{Antoniadis:1987dx}.  The term
$\phi h h$, on the other hand, provides an acceptable Higgs mixing term
at the electroweak scale.

If the symmetry of the model were just the one described above, however,
unwanted  terms could also be present. As a first example, we note that
the
singlet $\phi$ may form a mass term $\langle \phi \rangle F \bar{H}$.
In addition, the $SU(5)\times U(1)'$
gauge symmetry also allows trilinear terms
\ba
 F H h +  H\bar{f}\bar{h},
\ea
which give unacceptable mixings of ordinary fermions with the
colour-triplet fields and the Higgs multiplets. In order to avoid these
couplings, it
is  necessary  to impose a symmetry beyond $SU(5)\times U(1)'$. The
simplest possibility is to assume the $Z_2$ parity $H\ra -H$.   This
eliminates the terms (\ref{su5sup}), but cannot prevent the
mixing term $F \bar{H}\phi$.  One may invent another discrete
symmetry to prevent this term, while leaving other useful  terms
untouched, or $U(1)$ symmetries may be added to solve this problem.
In the present approach, as we have already discussed in the
Introduction, we would like to explore the possibility whether a
single $U(1)_A$ anomalous symmetry may answer all the above
questions in the context of the flipped  $SU(5) \times U(1)'$ model.   To
this
end, in the rest of this Section we describe our procedure,
and formulate the basic questions we think such
a symmetry should answer.

$\bullet$ Our first concern is the fermion mass textures.
To this end, we first study the constraints on the $U(1)$ charges. In the
MSSM case, in all fermion mass terms, the left- and right-handed fields
belong to different representations. Thus, the only constraints
on them arise from the anomaly cancellation conditions.  In
order to have an elegant structure, one may further demand
symmetric structures, but this demand restricts
considerably the acceptable fermion mass textures. Later, in the
final Section of this paper, we make some comments about this
restriction in the present class of models.

In the case of flipped $SU(5)\times U(1)'$, additional constraints
should be taken into account. The left- and right-handed down quarks
belong to the {\it same} representation, namely a decuplet of $SU(5)$,
whilst the lepton doublet is in the same pentaplet as the right-handed
up quark.


There is another interesting feature of models with
a unified or partially unified  gauge group. We recall first that the
fermion mass textures in the simple $U(1)$ extensions of the MSSM
are obtained  with only one singlet field, plus its conjugate,
via the following non-renormalizable terms:
\ba
\left(\frac{\langle\phi\rangle}{M_U}\right)^n  h q_i q_j,\,
\left(\frac{\langle\bar\phi\rangle}{M_U}\right)^n  h q_i q_j,
\ea
where $M_U$ is the unification scale and $q_i$ denotes a fermion field.
In the case of models with a larger gauge group, there are Higgs fields
$H,\bar{H}$ needed to break the symmetry down to that of the Standard
Model.
These Higgses may form an effective singlet combination $H\bar{H}$,
which can generate additional non-renormalizable contributions to the
fermion mass matrices.
This  creates a new hierarchical structure, in addition to the one
obtained from the singlets:
\ba
\left(\frac{\langle H\bar{H}\rangle}{M_U}\right)^m h q_i q_j.
\ea
The dominant term depends on the choice of the $H, {\bar H}$ charges,
and is also restricted by the following very important issue.

$\bullet$ We need  to deal with the extra triplets from the
pentaplets $h$ and $\bar{h}$. These should receive masses from the terms
$H H h + \bar{H} \bar{H} \bar{h}$, which combine them with the uneaten
components of $H,\bar{H}$, either at the 
renormalizable trilinear level or from some
higher-order non-renormalizable term. We have the freedom to select the
anomalous $U(1)$ charges of $H, \bar{H}$ so that these terms survive the
symmetry.

$\bullet$ In the old supersymmetric version of $SU(5)\times
U(1)'$~\cite{Antoniadis:1987dx}, a singlet $\phi_0$ with vev $\sim
m_W$ was introduced to deal with the Higgs mixing problem. This
adds more complications to the hierarchy problem. We have learned
from string theory that extra singlet vevs are determined through
the anomaly cancellation condition and are naturally of order
$10^{-1} M_{string}$, though this may be avoided by an acute
choice of superpotential. An alternative to a single singlet is to
choose the anomalous $U(1)$ charge assignments so that the Higgs
mixing term first appears as a high-order non-renormalizable term,
in which case a naturally small scale may appear: \ba {\cal
W}_{h\bar{h}} = \left(\frac{\langle\phi\rangle}{M_U}\right)^{r}
M_U h\bar{h} \ea gives a mixing of the right order for $r> 12$ to
$15$, with the precise value depending on the details of the model,
which fix the exact value of the ratio $\phi/M$ . We should note
that this is one of the most restrictive conditions on the
anomalous $U(1)_A$ charges. It will turn out that, in several cases,
this restriction results in a rather peculiar charge assignment.

\section{The origin of $R$-, $B$- and $L$-Violating Couplings
in Flipped $SU(5)\times U(1)'$}

In order to explore the constraints imposed on the $SU(5)\times U(1)'$
couplings by an Abelian non-anomalous symmetry $U(1)_A$, we
first need to identify the operators (\ref{sup2}) in  the
$SU(5)\times U(1)'$-invariant Yukawa superpotential.
Thus, we first search for possible higher-order non-renormalizable
gauge-invariant terms leading to the $R$-violating
couplings $\lambda$, $\lambda'$ and $\lambda''$.

The basic tree-level terms involve, in addition to the fermion mass
terms, only couplings of light with heavy states:
\ba
F F h &\ra
& q d^c h_d + d^c\nu^c D + q q D \label{pda} \\
F\bar{f}\bar{h}&\ra& q u^c h_u + \ell\nu^c h_u + q \bar{D} \ell +
d^c u^c \bar{D} \label{pdb}\\ \bar{f} h \ell^c &\ra& \ell e^c
h_d
\label{flh}
\ea
However, protons can decay through the
effective dimension-5 operator formed from the combination of the
couplings $q q D$ and $d^c u^c \bar{D}$, provided triplet mixing
occurs.
The invariant couplings providing masses to the triplets $D$ and $\bar D$
are:
\ba
H H h + \bar H \bar H \bar h
\ea
and possibly the bilinear combinations:
\be
\mu_H H\bar H + \mu h \bar h.
\ee
It is important to note that, in the minimal version of $SU(5)\times
U(1)'$, one triplet pair can be relatively light.
The reason~\cite{GRT} is that
the dimension-5 operators involving a pair of triplets are
proportional not only to the inverse of the mass, but also
to the cofactor of the corresponding element of the
triplet mass matrix. In the case of minimal $SU(5)\times U(1)$,
this means that proton decay is proportional to the
coefficient of the $H\bar H$ term $\mu_H$. If this term is absent or
suppressed, as is also required by the $F$-flatness conditions, the
associated operator is also suppressed, independently from the mass
of the triplets, which can be quite different.

In addition to the above tree-level terms, the $SU(5)\times U(1)'$
symmetry also allows trilinear couplings mixing
coloured triplets and Higgs doublets with fermion fields, of the forms:
\ba
F H h &\ra&  \langle \nu^c_H\rangle  d^c D  + h_d q d^c_H\label{FHh}\\
H \bar{f}\bar h &\ra &\langle \nu^c_H\rangle \ell h_u
\label{Hfbhb}
\ea
The  coupling (\ref{FHh}) induces a large mixing of $d^c$ with
$D$, whilst the term (\ref{Hfbhb}) gives unacceptably large
Higgsino-lepton mixing. An obvious discrete symmetry which may be
introduced to prevent these terms without affecting the terms in
(\ref{su5sup}) is $H\ra - H$.

We recall the absence of renormalizable $R$-violating terms due to
the GUT symmetry.
To identify higher-order $R$-violating terms, we search for all
possible invariant non-renormalizable couplings. At lowest (fourth)
order, we find
\ba
10_{1/2}\cdot 10_{1/2}\cdot 10_{1/2}'\cdot \bar{5}_{-3/2}
\ea
with two possible ways to make the contraction
\ba
either&[10_{1/2}\cdot 10_{1/2}] [10_{1/2}'\cdot \bar{5}_{-3/2}]\\
or &
[10_{1/2}\cdot 10_{1/2}'] [10_{1/2}\cdot \bar{5}_{-3/2}]
\ea
Two of the decuplets contain conventional families $F$, and one
is related to the $SU(5)$-breaking Higgs $H$ which,
at lowest order, gives the following couplings
\ba
\frac{H}{M} F F  \bar{f} &\ra&  \frac{\langle \nu^c_H\rangle}{M}
(\ell q d^c, u^c d^c d^c)\label{20}\\
&\ra & \lambda',\, \lambda''\nonumber
\ea
where $M$ is some high (unification or string) scale.
The second term is
\ba
\underline{10}_{1/2}'\cdot \bar{5}_{-3/2}\cdot \bar{5}_{-3/2}
\cdot 1_{5/2},
\ea
which corresponds to the fields
\ba
\frac H M \bar{f} \bar{f} \ell^c&\ra& \frac{\langle
\nu^c_H\rangle}{M}\ell \ell e^c\\
&\ra&\lambda\nonumber
\ea
At higher orders, we find the following dimension-5 operators which
violate baryon and lepton number:
\ba
\frac{\lambda_4^{ijkl}}{M_U}\ F_i F_j F_k \bar{f}_l,&
\displaystyle\frac{\lambda_5^{ijkl}}{M_U}\, F_i\bar{f}_j\bar{f}_k\ell_l^c
\label{dim5}
\ea
where the indices $i,j,k,l =1,2,3$ refer to the three generations.
In $SU(3)\times SU(2) \times U(1)$ notation, these are written
\ba
\frac{\lambda_4^{ijkl}}{M_U}\,q_iq_jq_k\ell_l,&
\displaystyle\frac{\lambda_5^{ijkl}}{M_U}\,u^c_iu^c_jd^c_k e^c_l
\ea
Although the induced amplitudes of dimension-5 operators are
suppressed
\footnote{We note that if there were universality for all
couplings $\lambda_4^{ijkl}$ and
$\lambda_5^{ijkl}$, the sum of all such operators would vanish
for symmetry reasons. However, there is no general argument that such a
universality assumption holds in string theory.}
compared to those arising from products of terms of the forms
(\ref{pda},\ref{pdb}), due to the fact that they arise as
non--renormalizable interactions scaled by some high mass scale, the
baryon-decay bounds on their
generalized Yukawa coupling constants are very restrictive. In
general, one has to impose $\lambda_4 <10^{-7}$ for operators
involving light quarks, whilst the constraints are less important
for $\lambda_5$.

\section{Anomaly Cancellation and Fermion Mass Textures in Flipped
$SU(5)\times U(1)'$.}

In this Section we explore the options in an $SU(5)\times U(1)'$ gauge
theory with representations  charged under an anomalous $U(1)_A$
symmetry. After finding successful charge assignments, and
requiring $SU(5)\times U(1)'$ invariance, we find the orders where
various mass terms appear in the Yukawa Lagrangian.

Our main concern here is the imposition of the anomaly
cancellation conditions. According to~\cite{Ibanez:1993fy}, in
order to obtain the canonical $\sin^2\theta_W$ value at the
unification scale, certain conditions should be satisfied by the
mixed anomalies of $U(1)_A$ with the gauge-group factors. In the
case of the Standard Model, we have three gauge groups and therefore the
Green-Schwarz~\cite{Green:1984sg} mechanism combined with the
$\sin^2\theta_W$ condition imposes two relations $A_2/A_1 =A_3/A_1
=5/3$. In the present case, since the effective field theory model
resulting from string is based on a higher gauge symmetry, we
should apply the anomaly cancellation conditions
at this level of symmetry. Thus the $U(1)_A$ charges should
be assigned directly at the GUT level, taking into account the
constraints imposed by the particular gauge symmetry.

Assume that the states have the generic $U(1)_A$ charge assignments shown
in Table~\ref{uxa}. We note that
the down-quark mass matrix is automatically  symmetric. Concerning
the up quark and charged lepton mass matrices, we have the freedom
to choose either symmetric or non-symmetric textures. We start
with the simplest case of symmetric textures.

\begin{table}
\begin{center}
\begin{tabular}{|c|c|}
\hline
State&Charge\\
\hline
$F_i\left({10},\frac{1}{2}\right)$&$\alpha_i$\\
$\bar f_i\left(\bar{5},-\frac{3}{2}\right)$&$\beta_i$\\
$e^c_i\left({1},\frac{5}{2}\right)$&$\gamma_i$\\
$H\left({10},\frac{1}{2}\right)$&$\delta$\\
$\bar H\left(\overline{10},-\frac{1}{2}\right)$&$\bar\delta$\\
$h\left({5},-1\right)$&$\epsilon$\\
$\bar h\left(\bar{5},1\right)$&$\bar \epsilon$\\
\hline
\end{tabular}
\end{center}
\caption{{\it Generic charge assignments under $U(1)_A$.}}
\label{uxa}
\end{table}

In this case, we obtain
\bea
\a{i}+\beta_j=\a{j}+\beta_i\label{csyma}\\
\beta_i+\gamma_j=\gamma_j+\beta_i
\label{csymb}
\eea
We further require that the  third-generation
$F F h, F\bar f \bar h, \bar f e^c h$  couplings appear in the
renormalizable trilinear superpotential, which implies the further
constraints
\begin{eqnarray}
\alpha_3+\beta_3+\bar\epsilon&=&0\nonumber\\
2\alpha_3+\epsilon&=&0
\label{ctree}\\
\beta_3+\gamma_3+\epsilon&=&0.
\nonumber
\end{eqnarray}
Equations (21)-(23) can be used to solve for $\beta_1,\beta_2,\beta_3,
 \gamma_1,
\gamma_2,\gamma_3$ and $ \e$ in terms of $\a1,\a2,\a3,\bar \e$. After imposing
the symmetric (\ref{csyma}),(\ref{csymb}) and trilinear
constraints (\ref{ctree}), the charge matrix  of the up, down quark
and lepton masses takes the form
\be
C^{FFh}=C^{F\fb\hb}=C^{\fb e^c h}=\left(\begin{array}{ccc}
2(\a1-\a3)&(\a1-\a3)+(\a2-\a3)&\a1-\a3\\
(\a1-\a3)+(\a2-\a3)&2(\a2-\a3)&\a2-\a3\\ (\a1-\a3)&(\a2-\a3)&0\\
\end{array}\right)\label{qcharge}
\ee
In order to have acceptable quark masses, we must impose
\be
\a1-\a3=\frac{n}{2}\ , \  \a2-\a3 =\frac{m}{2}\ \ \mbox{\rm
where}\ m+n\ne0,\ m,n=\pm1,\pm2,\dots .
\label{qi}
\ee
The resulting effective field theory model has to be anomaly-free,
and the introduction of an extra
$U(1)_A$ group factor leads to anomalies which should be cancelled.
The Green--Schwarz anomaly cancellation mechanism may cancel the pure
$U(1)_A$ and $U(1)_A$-gravitational anomalies, but
there are mixed anomalies of the form $A_i=\left(G_i G_i
U(1)_A\right)$, where $G_i=(SU(5),U(1)_Y)$. In terms of the
$U(1)_A$ charges, in the case of $SU(5)\times U(1)$ these are
written as
\begin{eqnarray}
A_5 &:& \sum\alpha_i+\frac{1}{2}\sum\beta_i+(\delta+\bar\delta)+
\frac{1}{2}(\epsilon+\bar\epsilon)\label{a5}\\ A_1 &:&
\frac{5}{2}\sum\alpha_i+\frac{45}{4}\sum\beta_i+\frac{25}{4}\sum\gamma_i+
\frac{5}{2}(\delta+\bar\delta)+5(\epsilon+\bar\epsilon)
\label{a1}
\end{eqnarray}
\begin{eqnarray}
A_0 &:& 5\sum \alpha_i^2-\frac{15}{2}\sum
\beta_i^2+\frac{5}{2}\sum\gamma_i^2
+5(\delta^2-{\bar\delta}^2)+5({\bar\epsilon}^2-\epsilon^2)=0\label{a0}
\end{eqnarray}
In the case of flipped $SU(5)\times U(1)'$,
gauge-coupling unification
imposes the following relation between the gauge couplings:
\bea
\frac{25}{\alpha_Y}=\frac{24}{\alpha_1}+\frac{1}{\alpha_5}.
\eea
Then, the standard value
\bea
\sin^2\theta_W&=&\frac{3}{8}
\eea
at the unification scale
leads to the following  condition for the anomalies
\be
\frac{A_1}{A_5}=10.
\label{A51}
\ee
In order to handle the above system of equations, we find it convenient
to change
to a new set of variables. We define $\delta_{\pm}=\delta\pm\bar\delta$
and express the $\a1,\a2$ charges in the anomaly condition equations
in terms of the integers $m,n$  from definitions (\ref{qi}).
Then, (\ref{A51}) can be solved in terms of
$\delta_{+}$:
\be
\delta_+=\frac{m+n}{3}+2\a3.
\label{e1}
\ee
Substituting in the quadratic constraint (\ref{a0}) and
solving for $\delta_{-}=\delta-\bar\delta$, we obtain the
expression
\be
\delta_{-}=-\frac{(\bar\epsilon+2\a3)(m+n+2\a3-\bar\epsilon)}{(m+n)/6+\a3},\
\;\a3\ne -\frac{m+n}{6}
\label{e2}
\ee
In this general
solution of the anomaly constraint equations, we have expressed all
the $U(1)_A$ charges of the various fields in terms of four
parameters, namely the integers $m,n$ and the charges $\alpha_3$
and $\bar\epsilon$. Alternatively, we may use (\ref{e2}) to
express one of $\alpha_3, \bar\epsilon$ in terms of $\delta_-$. In
Table~\ref{tab3}, we give all field charges in terms of the above free
parameters.

The above solution is valid as long as the denominator of the
expression defining $\delta_-$ in (\ref{e2}) is non-zero, i.e.,
whenever $\a3\ne -\frac{m+n}{6}$. In the particular case where
$\a3=-\frac{m+n}{6}$, we have no constraint for $\delta_{-}$, but
(\ref{a0}) gives the following two solutions for $\bar\epsilon$:
 \bea
\mbox{either}\;\;
\bar\epsilon&=&\frac{m+n}{3}
\label{sol2}\\
 \mbox{or}\;\;
  \bar\epsilon&=&\frac{2(m+n)}{3}.
\label{sol3}
\eea
Solution (\ref{sol3}) must, however, be rejected in the case of an
anomalous $U(1)$, since it leads to $A_5=A_1=0$. The
$U(1)_A$ charges of the acceptable solution are presented in
Table~\ref{tab4}.
\begin{table}
\centering
\begin{tabular}{|l|c|c|c|}
\hline field&\multicolumn{3}{c|}{generation}\\ \hline &1&2&3\\
\hline $F$&$\frac{n}{2}+\a3$&$\frac{m}{2}+\a3$&$\a3$\\ $\bar
f$&$\frac{n}{2}-\eb-\a3$&$\frac{m}{2}-\eb-\a3$&$-\eb-\a3$\\
$\ell^c$&$\frac{n}{2}+\eb+3\a3$&$\frac{m}{2}+\eb+3\a3$&$\eb+3\a3$\\
\hline \multicolumn{4}{|c|}{Higgs}\\ \hline $H$&$\delta$&$\bar
H$&$\bar\delta$\\ $h$&$-2\a3$&$\bar h$&$\eb$\\ \hline
\end{tabular}
\caption{ {\it The $U(1)_A$ charge assignments in solution (\ref{e2})
for symmetric mass matrices in flipped $SU(5)\times U(1)'$. The variables
$\delta,\bar\delta$ are not free parameters, but can be expressed
in terms of $\a3,\eb$ (see relations (\ref{e1},\ref{e2}).}}
\label{tab3}
\end{table}
\begin{table}
\centering
\begin{tabular}{|l|c|c|c|}
\hline field&\multicolumn{3}{c|}{generation}\\ \hline &1&2&3\\
\hline $F$&$\frac{2n-m}{6}$&$\frac{2m-n}{6}$&$-\frac{m+n}{6}$\\
$\bar f$&$\frac{2n-m}{6}$&$\frac{2m-n}{6}$&$-\frac{m+n}{6}$\\
$\ell^c$&$\frac{2n-m}{6}$&$\frac{2m-n}{6}$&$-\frac{m+n}{6}$\\
\hline \multicolumn{4}{|c|}{Higgs}\\ \hline $H$&$\delta$&$\bar
H$&$-\delta$\\ $h$&$\frac{m+n}{3}$&$\bar h$&$\frac{m+n}{3}$\\
\hline
\end{tabular}
\caption{{\it The $U(1)_A$ charge assignments in solution (\ref{sol2})
for symmetric mass matrices in flipped
 $SU(5)\times U(1)'$.}}
\label{tab4}
\end{table}

\section{Charged-Fermion Mass Matrices}

In the previous Section, we have shown that consistency with the
anomaly-cancellation conditions for the $U(1)_A$ charges leads to
two distinct solutions. It is interesting to observe that the
mass matrices depend in both solutions
only on the integer parameters $m,n$. The charge matrices now have
a universal form
\bea
C^{FFh}=C^{F\fb\hb}=C^{\fb e^c
h}=\left(\begin{array}{ccc} n&\frac{m+n}{2}&\frac{n}{2}\\
\frac{m+n}{2}&{m}&\frac{m}{2}\\ \frac{n}{2}&\frac{m}{2}&0\\
\end{array}\right)
\label{qna}
\label{qul}
\eea
It is important to note here that the quark and lepton
mass matrices have {\it identical} powers of non-renormalizable terms in
all entries, in contrast to the case of the MSSM~\cite{LR}. In
the MSSM, the quark  textures are also
expressed in terms of two parameters $m,n$ as above, but the
charged lepton mass matrix is given by
\bea
C^{\ell e^c
h}_{MSSM}=\left(\begin{array}{ccc} n&\frac{m+n}{2}&\frac{n}{2}\\
\frac{m+n}{2}&{m+p}&\frac{m+p}{2}\\ \frac{n}{2}&\frac{m+p}{2}&0\\
\end{array}\right),
\label{MssM}
\ea
where $p$ is another integer. Clearly, the higher GUT
symmetry in the present model is more restrictive than in the case
of the MSSM, and, as a result, this parameter is forced to
be zero: $p=0$.

The actual form of the mass matrices is obtained by legal
non-renormalizable terms involving powers of singlets which cancel
the charge of the corresponding entry in (\ref{qul}). As mentioned
above, there is a significant difference in this regard, as
compared to the MSSM case. In the MSSM, in the minimal case one
introduces a pair of singlet fields $\phi,\bar\phi$ which obtain
vevs and contribute to the mass matrix entries  through two
dimensionless parameters $\lambda=\frac{\phi}{M_U},
\bar\lambda=\frac{\bar\phi}{M_U}$. In  the present case, in
addition to $\phi, \bar\phi$ there is an effective singlet formed
by the combination  $\kappa =\frac{H\bar{H}}{M_U^2}$ of the Higgs
representations breaking the $SU(5)\times U(1)$ symmetry to that
of the SM. Therefore, the structure of the fermion mass textures
is much richer in this case. There are now three expansion
parameters entering in the mass matrices, namely
$\lambda,\bar\lambda$ and $\kappa$. All of them have vevs  with
well-defined magnitudes, as the first two are related to the
$D$-term cancellation, whilst the third one determines the
$SU(5)\times U(1)'$ breaking scale. Let $q_i,q_j^c$ denote
fermions of the generations $i,j$ respectively, and $h$ the Higgs
which couples to these fields. Then, a  particular mass entry with
$U(1)_A$ charge $C_{ij}$ has in general the form \ba \kappa^x
\lambda^y \bar\lambda^z q_i q_j^c h \ea which implies the
following condition on the related charges \bea C_{ij} + x
\delta_+ + y q_{\phi} + z \bar q_{\phi}=0, \label{NRterm} \eea
where $x,y,z$ are integers  representing the powers of the three
singlets required to make the term $U(1)_A$-invariant, and
$q_{\phi},\bar q_{\phi}$ the $U(1)_A$ charges of the singlets.
Without loss of generality, we may assume that
$q_{\phi}=-\bar q_{\phi}=1$, so the above relation can simply be
written as $C_{ij} + x\delta_++\omega=0$ with $\omega=y-z$. The
largest contribution arises when either of $y,z$ are zero, so we
may write the non-renormalizable term formally as follows: \ba
\kappa^x \rho^{|\omega|} q_i q_j^c h \ea where $$\rho=\lambda
\theta(\omega)+\bar\lambda\theta(-\omega)\label{rho}\,.$$

Before we analyze the mass matrices,
it is useful to make a rough estimate of the orders
of magnitude of the above parameters. The parameter $\kappa$ is
related to the $SU(5)\times U(1)'$ breaking scale $M_5$. In the most common
scenario, this scale is close to the MSSM unification scale, which
is  about two orders of magnitude below the string scale.  The
parameters $\lambda,\bar\lambda$ are related to the anomalous
$U(1)_A$ breaking scale. It is important to note here that one is
forced to take $\lambda\ne\bar\lambda$ since, in the case of an
anomalous $U(1)_A$ symmetry, we should  necessarily take $\phi\ne
\bar\phi$ in order to cancel the $D$ term. In particular,
the Green--Schwarz anomaly cancellation mechanism generates a
constant Fayet--Iliopoulos\cite{FI} contribution to the $D$ term
of the anomalous $U(1)_A$.  This is proportional to the trace of
the anomalous charge over all fields capable of obtaining
non--zero vevs. To preserve supersymmetry, the following
$D$--flatness condition should be satisfied~\cite{DSW,ADS}:
\ba
  \sum_{\phi_j} Q_j^X |\phi_j|^2&=&-\xi \ne 0,
\label{Deq}
\ea
where $ Q_j^X$ is the $U(1)_A$ charge of the singlet $\phi_j$. The
parameter $\xi$ depends on the common gauge coupling at
$M_{string}$, and on $M_{string}$ itself. In the present case, where
only two singlet fields $\phi,\bar\phi$ are involved, this
condition becomes
\ba
|\langle\bar\phi\rangle|^2- |\langle\phi\rangle|^2 &=&\xi.
\label{Dflat}
\ea
The scales of the $\phi_j$ and $\bar\phi_j$ vevs are naturally about
one order of magnitude smaller than $M_{string}$. Therefore, in a
natural scenario for the hierarchy of scales, we expect
\ba
\langle H\rangle \approx {\cal O} (10^{-1}) \langle\phi\rangle
\approx {\cal O} (10^{-2}) M_{string},
\ea
and subsequently $\kappa < \bar\lambda < 1$. As can be seen from
(\ref{Dflat}), the remaining parameter $\lambda$ is not completely
determined: due to the positivity of $\xi$, we need only impose
$\lambda < \bar\lambda$.

Up to now, we have imposed the gauge-invariance constraints on the
fermion mass entries in a manner analogous to the MSSM case. In
the present case, however, there are additional constraints
related to the appearance of the extra colour-triplet states
contained in $H,\bar{H}$ and $h,\bar{h}$ representations, which
should   be massive at some high scale. Indeed, triplet--doublet
splitting requires the existence of the couplings
$HHh$ and $\bar H\bar H \bar h$. If we want these couplings to be
present at some order $k$ and $\bar k$ respectively, we have to
impose two extra constraints:
\begin{eqnarray}
2\delta+\epsilon=k\\ 2\bar\delta+\bar\epsilon=\bar k
\end{eqnarray}
where $k, \bar k$ should be integers. We can now exchange the
parameters $\bar\epsilon, \a3$ of solution (\ref{tab3}) for the
integers $k, \bar k$:
 \bea
\eb&=&\frac{(9 k^2
+ 6 \kb^2 - 3\kb(m + n) - (m + n)^2 +
    k(15\kb - 4(m + n))}{3 (5 k+4 \kb-3 (m+n))}\\
    \a3&=&\frac{(6k^2 + 6\kb^2 - 14\kb(m + n) + 7((m + n)^2 +
    3k(4\kb - 5(m + n))}{6 (5 k+4 \kb-3 (m+n))}
\eea
for $(5 k+4 \kb-3 (m+n))\ne0$, while for the solution
(\ref{sol2}) we have
\bea
\bar k =\frac{2(m+n)}{3}-k\  ,\ \delta_{-}=k-\frac{m+n}{3}
\eea
These are the most general relations which correlate the values of
$\bar\epsilon$ and $\a3$ with the integer numbers $m,n,k,\bar{k}$.

The above solutions satisfy all our constraints, but we draw
attention to one subtle point. The anomaly $A_1$ should {\em not}
vanish, precisely because we are postulating an anomalous
$U(1)$. Indeed, putting the solutions into (\ref{a1}), we get \ba
A_1=10 A_5=-5(k-\bar k)\frac{k+\bar k-m-n}{5 k+4 \bar k-3 (m+n)}
\label{sl4} \ea for solution (\ref{e1}) and \ba A_1=10
A_5=\frac{10(m+n)}{3} \label{sl5} \ea for solution (\ref{e2}).

An important prediction can now be extracted from the last two
formulae for the coloured triplet couplings: we observe that the
solution (\ref{sl4})  involves the factor $k-\bar k$ in the
numerator. This tells us that, when
$k=\bar k$, the mixed anomalies vanish and the $U(1)$ symmetry
must be anomaly-free. Therefore, we infer that in the presence of
an anomalous $U(1)$ symmetry it is not possible to obtain both the
terms
$HHh$ and $\bar H\bar H\bar h$ at the same order. In particular, we
conclude that these terms {\it cannot provide simultaneously}
renormalizable trilinear superpotential masses for the two triplet pairs,
because it is not possible to have $k=\bar k =0$. This should not be
considered as a drawback, since, as was argued in Section 3, a light
triplet pair can be tolerated, because it does not
necessarily lead to rapid baryon decay.

It is worth pointing out here that one need not necessarily embed
this model in a usual string context. In this case, there is no necessity
to impose a non-anomalous $U(1)$ symmetry, and the mixed anomalies
may well vanish, provided that the $U(1)^3$ anomalies also vanish.
This may also happen to the present model, provided that we also include
additional pairs of states whose contribution to the
$U(1)_A^3$ anomaly exactly cancels the contributions of the standard
particles' $U(1)_A$ charges.
In this case, one can find other solutions which lead to a viable
low energy model, as we demonstrate below with a simple choice.

As an example of this particular case,
we work out here the simplest example where both
couplings appear in the trilinear superpotential~\footnote{We
note, however, that the phenomenology would be more interesting
if one of these couplings appears at higher order, resulting in a
light colored triplet.}. For $k=\bar k=0$, there exists only one
solution, with $\bar\epsilon=(m+n)/9$, $\a3 =-7 (m+n)/18$ and
$\delta_+=-\frac{4}{9}(m+n)$. We next seek solutions of (\ref{NRterm})
involving the minimum powers of the expansion parameters
$\kappa,\lambda,\bar\lambda$.
Bearing in mind the relative magnitudes of $\kappa,\bar\lambda$
estimated above, we may derive the possible lowest-order contributions to
the mass-matrix entries. The expansion parameter $\kappa$ has significant
contributions only for $x=0,1,2$, and one finds
\ba
11-\mbox{entry:}&x=0, \omega =-n;\, x=1, \omega=\frac{4 m-5 n}9;\,x=2,\omega=
\frac{8 m-n}9\nonumber\\
12-\mbox{entry:}&x=0, \omega =-\frac{m+n}2;\, x=1, \omega=-\frac{m+n}{18};\,
x=2,\omega=\frac{7(m+n)}{18}\nonumber\\
13-\mbox{entry:}&x=0, \omega =-\frac{n}2;\, x=1, \omega=\frac{8 m-n}{18};,\,
x=2,\omega=\frac{16 m+7 n}{18}\nonumber\\
22-\mbox{entry:}&x=0, \omega =-m;\, x=1,\omega=\frac{4n-5m}9;\,
x=2,\omega=\frac{8n-m}9\nonumber\\
23-\mbox{entry:}&x=0, \omega =-\frac m2;\, x=1,\omega=\frac{8n-m}{18};\,
x=2, \omega=\frac{7 m+16 n}{18}
\nonumber
\ea
We now present a specific example. Starting from the $\mu$ term, in order
to obtain $\mu\approx {\cal O}(m_W)$, we take
$\epsilon+\bar\epsilon \equiv \frac{8}{9}(m+n) = 16$, which implies
$m+n=18$.
Taking $m=4, n=14$, as a viable choice,  we solve  the equations (\ref{NRterm})
to determine the possible solutions for $x,y,z$.
Some comments are in order here.
It is convenient to replace $y,z$ with a new single variable $\omega=y-z$.
When a solution gives a positive (negative) value for $\omega$, this
means that
$\lambda (\bar\lambda)$ should appear in the operator. Taking into account this
and the hierarchy of scales discussed above, the following dominant
terms appear:
\bea
\frac{M^{FFh}}{\langle h\rangle}\approx\frac{M^{F\fb\hb}}{\langle \hb\rangle}
\approx\frac{M^{\fb e^c h}}{\langle h\rangle}\approx\left(\begin{array}{ccc}
\kappa^2\bar\l^6&\kappa^2\bar\l&\kappa^2\l\\
\kappa^2\bar\l&\bar\l^4&\bar\l^2\\
\kappa^2\l&\bar\l^2&1\\
\end{array}\right)
\label{Exaqul}
\eea
{}From (\ref{Dflat}) we may take $\lambda < \bar\lambda $, so that the
(13) entry becomes small. Since $\kappa <\bar\lambda$, we
conclude that a
correct hierarchical mass pattern can be obtained in a natural way.

We return now to the general case with a non-anomalous $U(1)$
symmetry. It is easy
to see that, among the possible $U(1)_A$ charge assignments,
there are numerous possibilities that could lead
to a viable fermion mass structure. The construction of a successful
effective field theory model is however
more complicated, as we have explained in previous sections,
since a large number of other
phenomenological constraints have to be respected. To start with,
we first calculate  the charges of the two bilinears $H\bar H$ and
$h \bar h$ in the solutions found above.
As has already been noted, they are both important,
since the first is directly related to
the fermion mass textures, whilst the second is related to the
$\mu$ term. The charges of the $H\bar H$ and $h\bar h$ terms are
\ba
\delta+\bar\delta=
\frac{2(k-\kb-m-n)(3(k+\kb)-2(m+n))}
{3(5 k+4 \bar k-3 (m+n))}
\\
\epsilon+\bar\epsilon\equiv r=
\frac{3k(k+\kb)+11(k+\kb)(m+n)-8(m+n)^2} {3(5 k+4 \bar k-3 (m+n))}
\ea for solution (\ref{e1}), and \ba \delta+\bar\delta=0 \ ,\
\epsilon+\bar\epsilon=\frac{2(m+n)}{3} \ea for solution
(\ref{sol2}). The second solution allows for a renormalizable
$H\bar H \phi$ term, which prevents the breaking of  $SU(5)\times
U(1)$ to the Standard Model. Thus we will elaborate further on
solution (\ref{sol2}).

In order to determine the appropriate $U(1)_A$
charges of the fields, we scan the above equations for
solutions with $\epsilon+\bar\epsilon={\rm integer} \ge
12$ at least. However, the following subtle point arises here when
we consider the sum $\delta+\bar\delta$: if
$\delta+\bar\delta$ is not an integer, then
$\left\langle H\bar H\right\rangle$ can fill in non--integer
entries in  the mass matrices. In this case, the parameters
$m,n$ involved may also take non-integer values, and therefore  we
have to reconsider our analysis. In the present construction we
consider only cases which admit integer $\delta_+$ values.

Our next concern is the $R$-violating terms whose implications
were discussed in Section 2. To estimate their effects in a
particular solution, we need to know at which order they appear.
This depends on the $U(1)_A$ charge each of these operators
carries, which has to be cancelled by the appropriate number of
singlets. In Tables~\ref{tabA3} and \ref{tabA4} we present the
$U(1)_A$ charges of the two tree-level couplings $H \bar f\bar h$
and $F_i{H}h$ for the two acceptable solutions found above.
Bearing in mind the limits from Table 1, we infer that, if the
above two couplings are to be present in the effective model, then
large values of $m,n$ and $\delta, k$ are needed in order to
suppress them sufficiently. What other possibilities might we
have? There are two other ways to avoid conflict with
phenomenological limits. (i) We note again that it is  possible to
avoid these terms by imposing a discrete symmetry on the Higgs
field: $H\ra - H$. This choice would have further consequences for
other types of operators. For example, the operator $H\bar H$
cannot appear with odd powers, unless the discrete symmetry
applies to both $H,\bar H\ra -H,-\bar H$. In this case, all even
and odd non-renormalizable terms involving $\kappa$ still remain,
and contribute to the mass matrices. (ii) A second possibility is
to consider solutions which predict non-integer values for the net
charges of the above operators. In this case, it is not possible
to form $U(1)_A$-invariant non-renormalizable terms, and the
operators do not appear in the effective theory. In the next
sections we present various solutions where each of the above
cases are realized.
\begin{table}
\centering
\begin{tabular}{|c|c|}
\hline
$F_1 H h$, $H \bar f_1 \bar h$&$\frac{k+n}{2}$\\
\hline
$F_2 H h$, $H \bar f_2 \bar h$&$\frac{k+m}{2}$\\
\hline
$F_3 H h$, $H \bar f_3 \bar h$&$\frac{k}{2}$\\
\hline
\end{tabular}
\caption{\it The ${U(1)}_A$ charges of the $F_i H h$ and $H \bar
f_i \bar h$ terms for the solution (\ref{tab3}).} \label{tabA3}
\end{table}
\begin{table}
\centering
\begin{tabular}{|c|c|}
\hline
$F_1 H h$, $H \bar f_1 \bar h$&$\delta +\frac{m+4n}{6}$\\
\hline
$F_2 H h$, $H \bar f_2 \bar h$&$\delta +\frac{4m+n}{6}$\\
\hline
$F_3 H h$, $H \bar f_3 \bar h$&$\delta+\frac{m+n}{6}$\\
\hline
\end{tabular}
\caption{\it The ${U(1)}_A$ charges of the $F_i H h$ and $H
\bar f_i \bar h$ terms for solution (\ref{sol2}).}
\label{tabA4}
\end{table}


\section{Neutrino Mass Textures and Proton-Decay Operators}

In this Section, we derive the mass textures for the neutral
fermion sector, i.e., the Dirac and Majorana mass matrices for the
left- and right-handed neutrinos. An interesting consequence
of the group structure and the anomalous $U(1)_A$ symmetry
that becomes clear immediately is the intimate relation of the
neutrino mass matrix structure to dimension-five proton
decay operators. We start with the neutrino physics.

In contrast to the minimal $SU(5)\times U(1)'$ model, an inevitable
consequence of extending the MSSM to a supersymmetric model with
flipped $SU(5)\times U(1)'$ symmetry is the appearance of non-zero
neutrino
masses. It is well known that recent neutrino data strongly suggest a
non-zero, although tiny, neutrino mass, so this
$SU(5)\times U(1)$ prediction  is very welcome. The
interesting fact in the present approach is that, once we have
fixed the $U(1)_A$ charges in order to rederive the observed
charged-fermion mass hierarchy, we make definite
predictions for the neutrino mass sector, as we now analyze
further. The
Dirac neutrino mass matrix is derived from the same couplings as that of
the up quarks, so its form is fixed from (\ref{qcharge}).
There are two different Majorana mass matrices, one for the left-
and another for the right-handed components, which have
vastly different mass scales. Direct left-handed Majorana masses may arise
from terms involving the combinations $\bar f_i \bar f_j$. The
general gauge invariant form of the relevant terms  is
 \ba
 m_{\nu}\bar{f}_i\bar{f}_j= \bar{f}_i\bar{f}_j \langle H\rangle^{10}
\frac{\langle h h \rangle}{M} \kappa^{2x} \rho^{|\omega|},
\ea
where $\kappa$ and $\rho$ are the expansion parameters introduced above,
$\omega$ is an integer, and $x$ a non-negative integer.
In order to compensate the $U(1)'$ charge, an additional vev
$\langle H\rangle^{10}$
is introduced, which  results in a big suppression
compared to the effective generation of light Majorana masses
through the see-saw mechanism. Let us point out here another
important difference from the MSSM case. In the MSSM, both terms
are of the same order, but in $SU(5)\times U(1)'$ the big suppression
of the $m_{\nu}$ contribution arises due to the particular charge
assignments of the representations under the
$U(1)'$ symmetry. In order to determine the effective
light-Majorana mass matrix, we further need the form of the heavy
right-handed neutrino mass texture. The gauge-invariant terms
contributing to the latter are
\ba
{\cal M}_{N} F_i F_J=F_i F_j \frac{\bar H\bar H}{M}
 \kappa^x\rho^{|\omega|}.
\label{HMM}
\ea
There are also possible higher-order contributions from invariants
of the type $F_i F_j H H H$, but these are expected to be
relatively suppressed.  Thus, this matrix is also completely
specified once we have made a specific choice of the
$U(1)_A$ charges.  Then, the effective light neutrino matrix
relevant to experimental detection is found to be
\ba
m_{\nu}^{eff}&=& m_{\nu}-\frac 14 m_D^T{\cal M}_{N}^{-1}
m_D\nonumber\\
&\approx&-\frac 14 m_D^T{\cal M}_{N}^{-1} m_D,
\label{meff}
\ea
since, according to the previous comment concerning the $\bar f_j \bar
f_i$ terms,
only the second contribution to $m_{eff}$ is relevant here. 

Hence,
in order to obtain the structure of the effective light Majorana
mass matrix, we need to know the structures of the matrices $m_D$
and ${\cal M}_N$. However, due to the GUT symmetry, the Dirac mass matrix
is already determined and has the same form as the down-quark mass
matrix.
In order to determine the heavy Majorana matrix (\ref{HMM}), we
need to calculate all possible combinations $C^{M}_{ij}=F_i
F_j(\bar H)^2$. Another subtle issue arises here, however:
if we wish to obtain non-zero heavy Majorana mass entries, the
charge matrix entries $C^{M}_{ij}$ have to be integers, which implies
that the charges of the quantities $F_i\bar H$ should be either
integers or half-integers. In the former case, the bilinears
$F_i\bar H$, which usually lead to unacceptable mixing, appear
in the superpotential. Thus it is desirable to obtain a
solution for the $U(1)_A$ charge assignments which provides
half-integer charges for the $F_i\bar H$ operators. We also note
that a  simple way to avoid these terms without projecting out the
useful heavy-neutrino contributions is to adopt the discrete
symmetry $\bar H \ra - \bar H$ as was also discussed previously.

In what follows, we make a general analysis and
present solutions  for both cases, without mentioning  further any
additional discrete symmetry. To express the charges of the
$F_i\bar H$ and $F_i F_j(\bar H)^2$ terms, we introduce the parameter
$r=\e+\eb$. Then the charges of the $F_i\bar H$ terms are
\be
{\rm charge}(F_i\bar H)=\frac{\kb-r}{2}+
\frac{1}{2}v_i
\ee
where $v_i$ is the $i$-th component of the vector
\be
{\bf v}={\left(\begin{array}{c}n\\ m\\ 0\\
\end{array}\right)}.
\label{vv}
\ee
We make  here an  important remark: as has been
explained on several occasions in this work, in order to have a
$\mu$ term, it is necessary to  ensure an integer value for the sum
of the doublet Higgs charges $r=\e+\eb$.

The above property has a
direct implication for the existence of the heavy Majorana mass
matrix. The latter is related to the charge entries
\be
{\rm charge}\left(F_i F_j(\bar H)^2\right)=\kb-r+ c_{ij}
\ee
which has just the structure of the fermion matrix entries,
shifted by the common value $\kb - r$. This implies that the heavy
Majorana mass matrix has the same texture, but with
its entries are all shifted by the power $\kb - r$. Thus, we end up
with a very precise form for the Majorana mass matrix, which gives 
definite
predictions for neutrino physics. If, in particular the scale
where the matrix
${\cal M}_N$ is formed is related to that of the charged lepton
mass matrix, then, at the unification scale we may write
\ba
{\cal
M}_N&=& \rho^{2 |r|}\frac{m_{\ell}}{\langle h\rangle} M_U,
\label{MN2}
\ea
where $m_{\ell}$ is the charged-lepton mass matrix, whose
form is dictated by the charge assignment (\ref{qcharge}).

In a similar way, we can examine whether a certain charge
assignment allows the existence of dangerous dimension-five operators:
we find that their charges are expressible in terms of the same
parameters.
The charges of the operators involving only matter fields  can be written
as follows:
\be
{\rm charge}\left(F_i F_j F_s {\bar f}_s\right)=
{\rm charge}\left(F_1 \fb_i \fb_j {\ell^c}_s\right)=
-r + c_{ij}+v_s,
\ee
where $c_{ij}$ is the charge of the $(ij)$ charge fermion entry in
(\ref{qcharge}), and $v_s$ is the $s$ component of the vector ${\bf
v}$ defined in (\ref{vv}). As has been shown earlier, the
first operator leads to the dangerous $QQQ\ell$ combination
which leads to fast proton decay. Operators of this type involving
the first two generations should have a rather small effective
coupling of the order
$\lambda_4\approx 10^{-6}$ or less.  Another set of dangerous $B$- and
$L$-violating  operators arises when we replace one of the matter
representations $F_i$ above with the Higgs field $H$ to obtain
$F_iF_jH\bar f_s$ and $H\bar f_i\bar f_j\ell^c_s$, respectively. In
this case, we find that their charges are simply the above ones
shifted by the amount $(k-v_s)/2$:
\be
{\rm charge}\left(H F_i F_j {\bar f}_s\right)= {\rm charge}\left(H
\fb_i \fb_j {\ell^c}_s\right)=-r+\frac{k}{2}+
c_{ij}+\frac{1}{2}v_s.
\ee
It turns out that these operators put rather stringent limits on
the $U(1)_A$ charge assignments. Indeed, as noted earlier in this paper,
these give $R$-violating couplings of the form
\be
\lambda_{ijs}''u^cd^cd^c\approx
 \left(\frac{\phi}{M}\right)^x\frac{\langle \nu_H^c\rangle}{M}u^cd^cd^c,
\ee
where $x$ is an appropriate power. We see from Table 1 that the
most  stringent bound comes  from the product of  couplings
 $\lambda_{112}'\lambda_{112}''$ which, in the case of flipped
$SU(5)$ are equal as can be seen from (\ref{20}). Here the
 exponent takes the form $x=|n-r+\frac{k+n}2|$ thus for an
expansion parameter $\rho\sim 0.2$, the bound $\lambda_{112}'
\lambda_{112}''\sim \lambda_{112}^{\prime 2} <
10^{-21}$ shown in Table~1 requires a rather large value of
$x\ge 14$. We also mention that the other two possibilities for
avoiding these operators are either to have non-integer charge, or to
introduce the
$R$-parity symmetry. Below, we present examples where all
there operators carry half-integer charges and therefore do not
appear in the Yukawa Lagrangian.

We have now all the ingredients needed for specific solutions
of the constraints. We present various examples with
reasonable charge assignments which lead to different types of
low-energy phenomenological models.

\section{Specific Examples}

There are numerous solutions for the anomalous $U(1)_A$ charges,
satisfying the anomaly cancellation conditions and the symmetric
mass matrix requirements we impose. In this Section our intention is
to show that there exist cases with natural sets of charges which
lead to viable models. Since, as has been shown, there are
numerous constraints that should be taken into account, we make a
systematic search of solutions with large values of $r$, so that
the $\mu$ term is sufficiently suppressed.

In Tables~\ref{ts1} to \ref{ts6} we present the solutions for the
$U(1)_A$ charges in five representative cases.
The solutions for the $U(1)_A$ charges in Tables~8, 9 and 13 give the
following general fermion mass texture at the unification scale:
\ba
\frac{m_{Q}}{\langle h\rangle},\;\frac{m_d}{\langle \bar h\rangle}
,\;\frac{m_{\ell}}{\langle \bar h\rangle}\propto\left(
\begin{array}{ccc} \rho^8&\rho^6&\rho^4\\
\rho^6&\rho^4&\rho^2\\\rho^4&\rho^2&1\\
\end{array}\right)
\label{sols12}
\ea
up to order one coefficients, where the parameter $\rho$ was
defined in (\ref{rho}). The case shown in Table~\ref{ts5} predicts the
following mass structure:
\ba
\frac{m_{Q}}{\langle h\rangle},\;\frac{m_d}{\langle \bar h\rangle}
,\;\frac{m_{\ell}}{\langle \bar h\rangle}\propto\left(
\begin{array}{ccc} \rho^6&0&\rho^3\\
0&\rho^3&0\\\rho^3&0&1\\
\end{array}\right).
\label{sols12v}
\ea
{}Finally, the solutions presented in Tables 11 and 12  predict the
texture
\ba
\frac{m_{Q}}{\langle h\rangle},\;\frac{m_d}{\langle \bar h\rangle}
,\;\frac{m_{\ell}}{\langle \bar h\rangle}\propto\left(
\begin{array}{ccc} \rho^{16}&\rho^6&\rho^8\\
\rho^6&\rho^4&\rho^2\\\rho^4&\rho^2&1\\
\end{array}\right).
\label{sols13v}
\ea
The heavy Majorana mass matrix is given for any of the above textures,
as has been shown in the previous Section, by
\ba
{\cal M}_N = \rho^{|r|} M_U \frac{m_{\ell}}{\langle \bar
h\rangle},
\label{sols12MN}
\ea
where values of the parameter $r$ are shown in the Tables.  All three
cases
have been shown to give a correct fermion mass hierarchy, so we
will not elaborate this point further~\footnote{For a complete
list of acceptable fermion mass textures see \cite{RRR}.
The second texture was firstly proposed in~\cite{Giudice}.
For a numerical
investigation of the first case, see\cite{LV} (for a non-string
version) and \cite{ELLN} (for a model of string origin).}.
Instead, we discuss the $B$- and $L$-violating operators and
the $\mu$ term.

To estimate the effects of the various $R$-violating
and $B$-violating operators, we present also in the Tables the
values of the various parameters $k,\bar k$, etc..
In the cases  $A,B$ and $C$, we show three examples with simple charge
assignments which give a natural fermion mass hierarchy. However,
in these three cases the $R$-violating terms  are not
sufficiently suppressed, so a further discrete
symmetry is needed to avoid them. Further, the
$\mu$ term is acceptable in cases $A,C$, but in case
$B$  it is not sufficiently suppressed, so the Higgs doublets
would have unacceptably large mixing, in the absence of any further
selection rule.

A solution to these problems requires more peculiar charge
assignments, which are shown in the three remaining examples we
present in this work. Solution D does not suppress sufficiently
the $\mu$ term, but it exhibits the interesting fact that most
of the $B$ and $L$-violating operators are suppressed, as
shown in Table~\ref{BLtable}.  Moreover, the $R$-violating
operators of the form
$HFF\bar f$ have half-integer charges, and therefore are not
present in the superpotential. In model $E$, the value
$r=11$ solves marginally the Higgs mixing problem. The
$B$-violating operators $FFF\bar f$ are also marginally suppressed,
their order of suppression being also shown in Table~\ref{BLtable}. As
$r$ increases, we observe that we find solutions which satisfy
both requirements, i.e., the $\mu$ term is of order $m_W$,
whilst baryon-decay operators are also sufficiently suppressed.
Solution $F$ is such an example: the value $r=13$ gives naturally a
value for  $\mu$ at the electroweak scale, whilst all the baryon-violating
operators of form
$FFF\bar f$ and $F\bar f\bar f \ell^c$ are highly suppressed, as
can be seen in the last column of Table~\ref{BLtable}. The
rather interesting consequence of this particular charge
assignment is the fact that all $R$-violating operators are
absent, as in cases $D$ and $E$.

\begin{table}
\centering
\begin{tabular}{|l|c|c|c|}
\hline field&\multicolumn{3}{c|}{generation}\\ \hline &1&2&3\\
\hline $F$&$1$&$3$&$5$\\ $\bar
f$&$-1$&$1$&$3$\\
$\ell^c$&$3$&$5$&$7$\\
\hline \multicolumn{4}{|c|}{Higgs}\\ \hline $H$&$1$&$\bar
H$&$5$\\ $h$&$-10$&$\bar h$&$-8$\\ \hline
\end{tabular}
\caption{\it Solution A: $k=-8,\bar k=2, m=-4, n=-8, r=-18$.}
\label{ts1}
\end{table}

\begin{table}
\centering
\begin{tabular}{|l|c|c|c|}
\hline field&\multicolumn{3}{c|}{generation}\\\hline &1&2&3\\
\hline $F$&$2$&$0$&$-2$\\ $\bar
f$&$2$&$0$&$-2$\\
$\ell^c$&$2$&$0$&$-2$\\
\hline \multicolumn{4}{|c|}{Higgs}\\ \hline $H$&$-\frac{3}{2}$&$\bar
H$&$\frac{3}{2}$\\ $h$&$4$&$\bar h$&$4$\\ \hline
\end{tabular}
\caption{\it Solution B: $k=1,\bar k=7, m=4, n=8, r=8$.}
\label{ts2}
\end{table}

\begin{table}
\centering
\begin{tabular}{|l|c|c|c|}
\hline field&\multicolumn{3}{c|}{generation}\\ \hline &1&2&3\\
\hline $F$&$-\frac{3}{2}$&$-3$&$-\frac{9}{2}$\\ $\bar
f$&$\frac{3}{2}$&$0$&$-\frac{3}{2}$\\
$\ell^c$&$-\frac{9}{2}$&$-6$&$-\frac{15}{2}$\\ \hline
\multicolumn{4}{|c|}{Higgs}\\ \hline $H$&$0$&$\bar H$&$-6$\\
$h$&$9$&$\bar h$&$6$\\ \hline
\end{tabular}
\caption{\it Solution C: $k=9,\bar k=-6, m=3, n=6, r=15$.}
\label{ts5}
\end{table}

\begin{table}
\centering
\begin{tabular}{|l|c|c|c|}
\hline field&\multicolumn{3}{c|}{generation}\\ \hline &1&2&3\\
\hline $F$&$5$&$-5$&$-3$\\ $\bar
f$&$8$&$-2$&$0$\\
$\ell^c$&$2$&$8$&$-6$\\
\hline \multicolumn{4}{|c|}{Higgs}\\
\hline $H$&$-\frac{11}2$&$\bar
H$&$\frac 72$\\ $h$&$6$&$\bar h$&$3$\\ \hline
\end{tabular}
\caption{\it Solution D: $k=-5,\bar k=10, m=-4, n=16$, $r=9$.}
\label{ts3}
\end{table}

\begin{table}
\centering
\begin{tabular}{|l|c|c|c|}
\hline field&\multicolumn{3}{c|}{generation}\\ \hline &1&2&3\\
\hline $F$&$3$&$-7$&$-5$\\ $\bar
f$&$12$&${2}$&$4$\\
$\ell^c$&$-6$&$-16$&$-14$\\
\hline \multicolumn{4}{|c|}{Higgs}\\ \hline $H$&$-\frac{9}{2}$&$\bar
H$&$-\frac{3}{2}$\\ $h$&$10$&$\bar h$&$1$\\ \hline
\end{tabular}
\caption{\it Solution E:$k=1,\bar k=2, m=-4, n=16$, $r=11$.}
\label{ts4}
\end{table}

\begin{table}
\centering
\begin{tabular}{|l|c|c|c|}
\hline field&\multicolumn{3}{c|}{generation}\\ \hline &1&2&3\\
\hline $F$&$-7$&$-5$&$-3$\\ $\bar
f$&$-8$&$-6$&$-4$\\
$\ell^c$&$-6$&$-4$&$-2$\\ \hline
\multicolumn{4}{|c|}{Higgs}\\ \hline $H$&$-\frac{15}2$&$\bar H$&$-\frac 52$\\
$h$&$6$&$\bar h$&$7$\\ \hline
\end{tabular}
\caption{\it Solution F: $k=-9,\bar k=2, m=-4, n=-8, r=13$.}
\label{ts6}
\end{table}


\begin{table}
\centering
\begin{tabular}{|l|l|c|c|c|}
\hline \multicolumn{2}{|c|}{Operator}& Model D& Model E & Model F\\
\hline
$q_1 q_1 q_2 \ell_1$ &$d^c_1 u^c_1 u^c_2 e^c_1$&13&  11 &-27  \\
\hline
$q_1 q_1 q_3 \ell_1$ &$d^c_1 u^c_1 u^c_3 e^c_1$&15& 13  & -25  \\
\hline
$q_1 q_2 q_3 \ell_1$ &
\begin{minipage}{2cm}
$d^c_1 u^c_2 u^c_3 e^c_1$\\
$d^c_2 u^c_1 u^c_3 e^c_1$\\
$d^c_3 u^c_1 u^c_2 e^c_1$
\end{minipage}
&5&3&$-23$ \\
\hline
$q_2 q_1 q_2 \ell_1$ &$d^c_2 u^c_1 u^c_2 e^c_1$& 3&1 & $-25$    \\
\hline
$q_2 q_2 q_3 \ell_1$ &$d^c_2 u^c_2 u^c_3 e^c_1$& -5& -7&-21  \\
\hline
$q_3 q_1 q_3 \ell_1$ &$d^c_3 u^c_1 u^c_3 e^c_1$& 7&  4 & $-21$    \\
\hline
$q_3 q_2 q_3 \ell_1$ &$d^c_3 u^c_2 u^c_3 e^c_1$& -3& -5 & $-19$\\
\hline
$q_1 q_1 q_2 \ell_2$ & $d^c_1 u^c_1 u^c_2 e^c_2$&3& 1&$-25$\\
\hline
$q_1 q_1 q_3 \ell_2$ & $d^c_1 u^c_1 u^c_3 e^c_2$&5&$3$&$-23$\\
\hline
$q_1 q_2 q_3 \ell_2$ &
\begin{minipage}{2cm}
$d^c_1 u^c_2 u^c_3 e^c_2$\\
$d^c_2 u^c_1 u^c_3 e^c_2$\\
$d^c_3 u^c_1 u^c_2 e^c_2$
\end{minipage}
 &-5&-7&$-21$ \\
\hline
$q_2 q_1 q_2 \ell_2$&$d^c_2 u^c_1 u^c_2 e^c_2$&-7&$-9$ &$-23$\\
\hline
$q_2 q_2 q_3 \ell_2$ &$d^c_2 u^c_2 u^c_3 e^c_2$&
-15&$-17$&$-19$\\
\hline
$q_3 q_1 q_3 \ell_2$ &$d^c_3 u^c_1 u^c_3 e^c_2$&-3&$-5$&-19\\
\hline
$q_3 q_2 q_3 \ell_2$ &$d^c_3 u^c_2 u^c_3 e^c_2$&-13 &-15&$-17$\\
\hline
$q_1 q_1 q_2 \ell_3$ &$d^c_1 u^c_1 u^c_2 e^c_3$&$5$&3&$-23$\\
\hline
$q_1 q_1 q_3 \ell_3$ &$d^c_1 u^c_1 u^c_3 e^c_3$&7&$5$ &$-21$\\
\hline
$q_1 q_2 q_3 \ell_3$ &
\begin{minipage}{2cm}
$d^c_1 u^c_2 u^c_3 e^c_3$\\
$d^c_2 u^c_1 u^c_3 e^c_3$\\
$d^c_3 u^c_1 u^c_2 e^c_3$
\end{minipage}
&-3&-5&$-19$\\
\hline
$q_2 q_1 q_2 \ell_3$ &$d^c_2 u^c_1 u^c_2 e^c_3$&-5 &$-7$&$-21$\\
\hline
$q_2 q_2 q_3 \ell_3$ &$d^c_2 u^c_2 u^c_3 e^c_3$&-13&$-15$ &$-17$\\
\hline
$q_3 q_1 q_3 \ell_3$ &$d^c_3 u^c_1 u^c_3 e^c_3$&-1& $-3$&$-17$\\
\hline
$q_3 q_2 q_3 \ell_3$ &$d^c_3 u^c_2 u^c_3 e^c_3$&-11& -13&$-15$\\
\hline
\end{tabular}
\caption{\label{BLtable} {\it Dimension-5  proton
decay operators originating from the $SU(5)\times U(1)'$ invariants
$FFF\bar f$ and $F \bar f\bar f \ell^c$ are presented in the first and second
columns. In columns 3 to 5, we present the
numerical values of the ${U(1)}_A$ charges (and therefore the powers of
the corresponding non-renormalizable terms)
for models $D,E$ and $F$ respectively.}}
\end{table}

\section{Conclusions}

The appearance of anomalous
Abelian $U(1)_A$ symmetries is a generic phenomenon
in string constructions. They act as family symmetries
and, together with the gauge symmetry and other discrete (string)
symmetries, determine the possible forms of the superpotential couplings
of fermions.
In this paper we have analysed in detail the implications of a
general $U(1)_A$ anomalous symmetry in the context of the
flipped $SU(5)\times U(1)'$ model derivable
from string.

We imposed the appropriate anomaly cancellation
conditions  on the
anomalous Abelian charges of the $SU(5)\times U(1)'$
superfields, so as to obtain the canonical value
for the weak mixing angle: $\sin^2\theta_W=3/8$ at the unification scale.
For simplicity, we have restricted our attention to symmetric mass
textures,
although non-symmetric mass matrices for the up quarks and the
leptons are also a viable possibility.
We further  used the known bounds from
low-energy experimental physics on $L$- and $B$-violating Yukawa
couplings, as well as the acceptable
ranges of the mixing angles and the
fermion masses, to constrain the possible $U(1)_A$ charges
of the matter and Higgs representations of $SU(5)\times U(1)'$.

We have provided solutions of the anomaly-cancellation constraints
and showed that we can build  low-energy effective models
providing the correct fermion mass spectrum with rather simple
charge assignments.  The success of simple, small $U(1)_A$ charges
is spoiled, however, by the appearance of $R$-violating
interactions, as well as $B$-violating decays at unacceptable
levels and a large mixing term $\mu$ for the ordinary Higgs
doublets. Nevertheless, we find models  which are compatible with
all the known experimental facts, albeit with relatively large
values for the $U(1)_A$ charges. The $R$-violating terms are
prohibited, and baryon- and lepton-violating operators are highly
suppressed.

In the present work, we have shown how the anomalous $U(1)_A$
in string-derivable models based on  intermediate gauge symmetries
may contribute in the construction of a phenomenologically successful
theory. Although we have in mind string-derived models, in order to
make our analysis as general as possible,  we have
incorporated only  the most common features of present-day string
constructions. We have relaxed several of the simplified
assumptions  often used in the recent literature, but,
at this stage of our approach, we have retained some which
may not give a fully realistic picture. In the rest of this Section,
we discuss briefly these points and the limitations
of our procedure, and make a few comments on possible future extensions of
our analysis.

A crucial point for the consistency of the above symmetric scheme
is that the expansion parameter $\l$ in the up-quark sector differs
from the corresponding parameter $\epsilon$ in the down-quark
sector. In the context of the MSSM, $\l =\phi/M_1$ and $\e =\phi/M_2$,
where  $M_1$ and $M_2$ are  related to the scales where the up and down
quark mass matrices are formed. The
viability of most phenomenological explorations of
$U(1)$ family-symmetry models is based on the
observation~\cite{Ibanez:1994ig}
that it is  possible for additional vector-like Higgs pairs to
acquire their mass via spontaneous breaking after
compactification. Then, a strong  violation of the $SU(2)_R$
symmetry of the quark sector may occur, and as a result
$M_1\ne M_2$~\cite{Ibanez:1994ig}. This provides the possibility
that $\l \ne \e$, and a correct mass hierarchy may result.

In usual compactification scenarios, however,
including also the cases of the free-fermionic string models, this
is not the case. In fact, there is only one scale: $M_1=M_2\equiv
M_{string}$, and therefore the above two parameters coincide:
$\lambda=\epsilon$ and consequently $\bar\lambda =\bar\epsilon$.
Hence, if we are not able to distinguish the up- and
down-quark expansion mass parameters, it is impossible to obtain a
correct hierarchical mass pattern.

In realistic string scenarios, the discrimination between the
various fermion mass matrices is based on completely different
observations. There are basically three sources of the breaking of the
up-/down-quark mass matrix symmetry.  First, there are many
singlet fields acquiring vevs which couple differently to the
various types of quarks. Secondly, there are multiple $U(1)$
symmetries, although  only one is anomalous. Finally, fermion fields
are not always charged `symmetrically', eventually leading  to
non-symmetric mass matrices  whenever the non-Abelian structure of
the model makes this possible.
In the present case of the $SU(5)\times U(1)$ model, only the
down-quark mass matrix is always  symmetric, the reason being that
the left-handed $SU(2)$-doublet quarks and the
right-handed $SU(2$-singlet down
quarks belong to the same $SU(5)\times
U(1)$ representations. On the other hand, the up-quark and charged-lepton
mass matrices need not be symmetric.

A question one should then answer is: what are
the minimal modifications of the above scenario leading to
a natural hierarchical mass spectrum?  If our aim is to obtain an
economical way of constructing viable mass matrices, we should
avoid introducing a large number of singlets or symmetries. Thus,
in more realistic cases, in order to emulate a  realistic string
scenario, one should relax the  symmetry constraints on the
$U(1)$ charges of the up-quark mass matrix. On the other hand,
one may  retain the
same number of singlets and employ a single anomalous $U(1)$
family symmetry, if this is sufficient
to obtain a satisfactory result.
We plan to analyse this scenario in a future publication.

\end{document}